\documentclass[12pt]{article}
\usepackage{amssymb}

\usepackage{graphicx}
\usepackage{amsmath}

\begin{document}

\title{CLASSICAL AND QUANTIZED ASPECTS OF DYNAMICS IN FIVE-DIMENSIONAL
RELATIVITY}
\date{}
\author{}
\maketitle


\begin{center}
Paul S. Wesson$^{1,\,2}$
\end{center}

\begin{enumerate}
\item Dept. Physics, Univ. of Waterloo, Waterloo, ON N2L 2G1, Canada
\linebreak Keywords: particle motion, quantization, uncertainty, induced
matter, membranes \ \ \ \ \ \ \ \ \ \ \ \ \ \ \ \ \ \ \ \ \ \ \ \ \ \ \ \ \
\ \ \ \ \ \ \ \ \ PACs: 04.50.+h: 04.20.Cv

\item Correspondence: Mail = (1) above; fax=(519) 746-8115;

email=wesson@astro.uwaterloo.ca. \pagebreak
\end{enumerate}

\begin{center}
\bigskip {\LARGE CLASSICAL AND QUANTIZED ASPECTS OF DYNAMICS IN
FIVE-DIMENSIONAL RELATIVITY}
\end{center}

\bigskip

\underline{Abstract}

\qquad A null path in 5D can appear as a timelike path in 4D, and for a
certain gauge in 5D the motion of a massive particle in 4D obeys the usual
quantization rule with an uncertainty-type relation. \ Generalizations of
this result are discussed in regard to induced-matter and membrane theory.

\bigskip

\section{\protect\underline{Introduction}}

Recently there have been two related results in dynamics from
five-dimensional relativity which are remarkable. (a) When a theory like
general relativity is extended from 4D to 5D a fifth force appears, and this
can manifest itself in spacetime [1-16]. In 4D it is well known that the
force (per unit inertial mass) and velocity are constrained by the
orthogonality condition $f_{\alpha }u^{\alpha }=0\left( \alpha
=0,\,123\right) $. In 5D the corresponding condition is $f_{A}\,u^{A}=0%
\left( A=0,\,123,4\right) ,$ but then $f_{\alpha }\;u^{\alpha
}=-f_{4}\;u^{4}\neq 0$ [7]. The fifth force does not manifest itself for a
certain class of metrics based on pure canonical coordinates [2], for
metrics where the coordinates can be chosen so as to make the velocity in
the extra dimension comoving [8], and for metrics parametized in such a way
as to make it disappear [15]. In general, however, the fifth force exists
for 5D metrics which depend on the extra coordinate $x^{4}=l$, and is
therefore present in both induced-matter theory [2-8, 12, 13, 16] and brane
theory [9-11, 14, 15]. The fifth force is different from others in 4D
dynamics in that it acts \underline{parallel} to the 4-velocity $u^{\alpha }$%
, so it is natural to express its effects in terms of the momenta or the
(inertial) rest mass $m$ of the particle that feels it [2, 6, 8, 9]. This is
why $m$ can be related to $l$ or its rate of change, depending on the
coordinates, as will be discussed below. \ (b) When a manifold is extended
from 4D to 5D the spacetime line element $ds$ is embedded in a larger line
element $dS$, and particles which are massive and move on timelike paths in
4D with $ds^{2}>0$ can move on null paths in 5D with $dS^{2}=0$ [16-20]. \
Conventional causality is defined by $ds^{2}\geq 0$, but this is compatible
with $ds^{2}\geq 0$ or $ds^{2}\leq 0$ [18]. Photons move on 5D geodesics
with $ds^{2}=0=dS^{2}$, but massive particles can also move on geodesics
with $dS^{2}=0$ provided 4D paths are allowed with $m=m\left( s\right) $. \
This will in general be the case if the fifth force acts, as outlined above.
The mass variation does not manifest itself if the parameter along the path
is specially chosen to make it disappear [19]; but to make contact with
standard \ 4D dynamics it is logical to use $s$ as the parameter, so in
general $m=m\left( s\right) $ in both induced-matter theory and brane
theory. The null condition $dS^{2}=0$ is compatible with the conventional
relation between the energy, momentum and mass of a particle in 4D but means
that its ``energy'' is zero in 5D. This agrees with the fact that Campbell's
theorem guarantees the local embedding of any 4D Riemannian space which is
curved and contains matter in a 5D Riemannian space which is Ricci-flat and
empty [21-24]. The precise form for the energy of a particle in 4D depends
on the 5D metric or the coordinates, as will be discussed below.

The results summarized in (a), (b) above are startling but classical in
nature. We wish to compliment them below by deriving some results related to
the quantized aspects of particle dynamics. It is already apparent that 5D
physics is enriched by the addition of an extra dimension but perforce
encumbered by the need to make a choice of 5D coordinates (or gauge) which
gives back recognizeable 4D physics. Obviously the group of 5D coordinate
transformations $x^{A}\rightarrow \overline{x}^{A}\left( x^{B}\right) $ is
wider than the group of 4D coordinate transformations $x^{\alpha
}\rightarrow \overline{x}^{\alpha }\left( x^{\beta }\right) $, and wider
than the restricted group of transformations (diffeomorphisms) $x^{\alpha
}\rightarrow \overline{x}^{\alpha }\left( x^{\beta }\right) $, $%
x^{4}\rightarrow \overline{x}^{4}\left( x^{4}\right) $ sometimes used. In
the next section, we will therefore start without apology from a metric
which is chosen (with hindsight) to give back recognizeable 4D physics. This
metric does not look like the canonical metric of induced-matter theory
(which is basically a factor in $l^{2}$ multiplied onto an Einstein metric
plus an extra flat part) or the warp metric of brane theory (which is
basically an exponential factor in $l$ multiplied onto an Einstein metric
plus an extra flat part). After deriving some results from the new metric we
will, however, reinterpret it and put it into context. Our results will
support the conjecture [8] that classical 5D physics can lead to quantized
4D physics.

\section{\protect\underline{The 5D Planck and Einstein Gauges}}

The 5D line element given by $dS^{2}=g_{AB}dx^{A}dx^{B}\left(
A=0,\,123,\,4\right) $ contains the 4D one $ds^{2}=g_{\alpha \beta
\,}dx^{\alpha }dx^{\beta }\left( \alpha =0,\,123\right) $. We can use 4 of
the available 5 degrees of coordinate freedom to set $g_{4\alpha }=0$, which
in the old Kaluza-Klein theory were identified with the electromagnetic
potentials. We could use the remaining degree of freedom to set $\left|
g_{44}\right| =1$ and thereby supress the scalar potential, but in some
newer versions of Kaluza-Klein theory this is related to the Higgs potential
[8], and it proves more instructive to restrict this only via $%
g_{44}=g_{44}\left( x^{4}\right) $ without at the outset carrying out the
coordinate transformation that would make it constant. (We will do this
later; the 5D geodesic equation for unrestricted $g_{AB}$ is considered in
reference 3.) The problem is so far general in a mathematical sense if we
allow the spacetime metric to be $g_{\alpha \beta }=g_{\alpha \beta }\left(
x^{\gamma },\,x^{4}\right) $. Exact solutions of the field equations are
known that have this property, including cosmological ones which reduce on
the hypersurfaces $x^{4}=$ constants to Friedmann-Robertson-Walker models
and agree with observations [8]. However, the focus here is on particle
dynamics, and since observations indicate that there is no explicit
incursion of the fifth dimension into local spacetime, we put $g_{\alpha
\beta }=g_{\alpha \beta }\left( x^{\gamma }\right) $, which means that the
Weak Equivalence Principle is a symmetry of the 4D part of the 5D metric. We
label the time, space and extra coordinates by $x^{0}=ct,\,x^{123}$ and $%
x^{4}=l$, taking them all to have physical dimensions of length. To
distinguish physically between our starting gauge and a later one, it is
useful to retain conventional dimensions for the speed of light $c$,
Planck's constant $h$ and the gravitational constant $G$. We refrain at the
outset from physically identifying $x^{4}=l$, though it is apparent from the
comments in Section 1 that we expect it to be related to the (inertial) rest
mass of a test particle. We also expect $m=m\left( s\right) $. It should be
noted that this does not violate the usual condition $g_{\alpha \beta
}u^{\alpha }u^{\beta }=1$ for the 4-velocities $u^{\alpha }\equiv dx^{\alpha
}\diagup ds$. This is a normalization condition on the velocities, not a
coordinate condition on the metric, so we can adopt it. Multiplying this
condition by $m^{2}$ gives $p^{\alpha }p_{\alpha }=m^{2}$ where $p^{\alpha
}\equiv mu^{\alpha }$ with \underline{no restriction} on $m=m\left( s\right)
$. In other words, the (squares of the) energy $E^{2}=m^{2}c^{4}u^{0}u_{0}$
and 3-momentum $p^{2}=m^{2}c^{2}\left(
u^{1}u_{1}+u^{2}u_{2}+u^{3}u_{3}\right) $ satisfy $%
E^{2}-p^{2}c^{2}-m^{2}c^{4}=0$ even if the mass varies in spacetime. The
last-noted relation is closely obeyed by real particles; but while the
opposite sign for $g_{44}$ has been taken in so-called two-time metric
[25-30 and below], the implication is that our 5D metric should\ have [+(- -
-)-] for its signature.

Based on the preceding, consider a line element given by
\begin{equation}
dS^{2}=\frac{L^{2}}{l^{2}}\;g_{\alpha \beta }\left( x^{\gamma }\right)
dx^{\alpha }dx^{\beta }-\frac{L^{4}}{l^{4}}dl^{2}\;\;\;\;\;.
\end{equation}
Here $L$ is a constant length introduced for dimensional consistency whose
physical meaning we will return to below. The Lagragian density $\mathcal{L\,%
}=\left( dS\diagup ds\right) ^{2}$ [19] has associated with it 5-momenta
given by
\begin{subequations}
\begin{eqnarray}
P_{\alpha } &=&\frac{\partial \mathcal{L}}{\partial \left( dx^{\alpha
}\diagup ds\right) }=\frac{2L^{2}}{l^{2}}g_{\alpha \beta }\frac{dx^{\beta }}{%
ds} \\[0.25in]
P_{l} &=&\frac{\partial \mathcal{L}}{\partial \left( dl\diagup ds\right) }=-%
\frac{2L^{4}}{l^{4}}\frac{dl}{ds}\;\;.
\end{eqnarray}
These define a 5D scalar which is the analog of the one used in 4D quantum
mechanics:
\end{subequations}
\begin{eqnarray}
\int P_{A}\,dx^{A} &=&\int \left( P_{\alpha }dx^{\alpha }+P_{l}dl\right)
\notag \\
&=&\int \frac{2L^{2}}{l^{2}}\left[ 1-\left( \frac{L}{l}\,\frac{dl}{ds}%
\right) ^{2}\right] ds\;\;.
\end{eqnarray}
This is zero for $dS^{2}=0$, since then (1) gives
\begin{equation}
l=l_{0}e^{\pm s/L},\;\;\;\;\frac{dl}{ds}=\pm \frac{l}{L},
\end{equation}
where $l_{0}$ is a constant. The second member of this shows why some
workers have related the (inertial) rest mass of a particle to $l$ [8] and
some to its rate of change [20] with consistent results: the two
parametizations are essentially equivalent. In both cases, the variation is
slow if $s\diagup L\ll 1$ (see below). We prefer to proceed with the former,
because it is simpler. Also, this makes the first part of the 5D line
element in (1) essentially the element of the usual 4D action $mcds$. It
should be noted in passing that in forming the total action from the latter
quantity, the $m$ should go inside the integral, even in 4D theory [31]. The
appropriate parametization with the problem as set up here is $l=h/mc$, the
Compton wavelength of the particle. The latter has finite energy in 4D, but
zero ``energy'' in 5D because $\int P_{A}dx^{A}=0$.

The corresponding quantity in 4D is $\int p_{\alpha }dx^{\alpha }$ and for a
massive particle is nonzero. Using relations from the preceding paragraph,
it is
\begin{equation}
\int p_{\alpha }dx^{\alpha }=\int mu_{\alpha }dx^{\alpha }=\int \frac{hds}{cl%
}=\pm \frac{h}{c}\,\frac{L}{l}.
\end{equation}
The fact that this can be positive or negative goes back to (4), but since
the motion is reversible we will suppress the sign in what follows for
convenience. We will also put $L/l=n$, anticipating a physical
interpretation which indicates that it is not only dimensionless but may be
a rational number. Then (5) says
\begin{equation}
\int mcds=nh.
\end{equation}
Thus the conventional action of particle physics in 4D follows from a null
line element (1) in 5D.

The other scalar quantity that is of interest in this approach is $%
dp_{\alpha }dx^{\alpha }$. \ (It should be recalled that $dx^{\alpha }$
transforms as a tensor but $x^{\alpha }$ does not.) Following the same
procedure as above there comes

\begin{equation}
dp_{\alpha }dx^{\alpha }=\frac{h}{c}\left( \frac{du_{\alpha }}{ds}\,\frac{%
dx^{\alpha }}{ds}-\frac{1}{l}\,\frac{dl}{ds}\right) \frac{ds^{2}}{l}\;\;.
\end{equation}
The first term inside the parenthesis here is zero if the acceleration is
zero or if the scalar product with the velocity is zero as in conventional
4D dynamics (see Section 1). But even so, there is a contribution from the
second term inside the parenthesis which is due to the change in mass of the
particle. This anomalous contribution has magnitude
\begin{equation}
\left| dp_{\alpha }dx^{\alpha }\right| =\frac{h}{c}\left| \frac{dl}{ds}%
\right| \frac{ds^{2}}{l^{2}}=\frac{h}{c}\frac{ds^{2}}{Ll}=n\frac{h}{c}\left(
\frac{dl}{l}\right) ^{2},
\end{equation}
where we have used (4) and $n=L\diagup l$. The latter implies $dn\diagup
n=-dl\diagup l=dK_{l}\diagup K_{l}$ where $K_{l}\equiv 1\diagup l$ is the
wavenumber for the extra dimension. Clearly (8) is a Heisenberg-type
relation, and can be written
\begin{equation}
\left| dp_{\alpha }dx^{\alpha }\right| =\frac{h}{c}\frac{dn^{2}}{n}\;\;.
\end{equation}
This requires some interpretation, however. Looking back at the 5D line
element (1), it is apparent that $L$ is a length scale not only for the
extra dimension but also for the 4D part of the manifold. (There may be
other scales associated with the sources for the potentials that figure in $%
g_{\alpha \beta }$, and these may define a scale via the 4D Ricci scalar $R$%
, but we expect that the 5D field equations will relate $R$ to $L$, as will
be illustrated below.) As the particle moves in spacetime, it therefore
``feels'' $L$, and this is reflected in the behaviour of its mass and
momentum. Relations (6) and (9) quantify this. If the particle is viewed as
a wave, its 4-momenta are defined by the de Broglie wavelengths and its mass
is defined by the Compton wavelength. The relation $dS^{2}=0$ for (1) is
equivalent to $P_{A}P^{A}=0$ or $K_{A}K^{A}=0.$ The question then arises of
whether the waves concerned are propagating in an open topology or trapped
in a closed topology. In the former case, the wavelength is not constrained
by the geometry, and low-mass particles can have large Compton wavelengths $%
l=h\diagup mc$ with $l>L$ and $n$ $=L\diagup l<1$. In the latter case, the
wavelength cannot exceed the confining size of the geometry, and high-mass
particles have small Compton wavelengths with $l\leq L$ and $n\geq 1$. By
(9), the former case obeys the conventional Uncertainty Principle while the
latter case violates it. This subject clearly needs an in-depth study, but
with the approach adopted here we tentatively identify the former case as
applying to real particles and the latter case as applying to virtual
particles.

The fundamental mode $\left( n=1\right) $ deserves special comment. This can
be studied using (6)-(9), or directly from (1) by using $l=h\diagup mc$ with
$dS^{2}=0$. The latter procedure gives $\left| dm\right| =mds\diagup L$
which with (6) yields $m=\left( \int mcds\right) \diagup cL=nh\diagup cL$.
This defines for $n=1$ a fundamental unit mass, $m_{0}=h\diagup cL$. In
general $L$ is a scale set by the problem, analogous to the ``box'' size in
old wave mechanics. In cosmology, we expect $L$ to be related to the
cosmological constant $\Lambda $. This inference is backed by detailed
analysis of the field equations and an examination of certain exact
solutions thereof [2, 5, 8, 9 and below]. These give $\Lambda =3\diagup
L^{2} $. The cosmological value of $L=\left( 3\diagup \Lambda \right) ^{1/2}$
is a maximum for this parameter, defining a minimum for $m_{0}$ that applies
even to particle physics. Astrophysical data indicate a positive value for $%
\Lambda $ of approximately $3\times 10^{-56}$cm$^{-2}$, though in view of
observational uncertainties this should be taken as a constraint rather than
a determination [32-36]. The noted value corresponds to a density for the
vacuum in general relativity of $\Lambda c^{2}\diagup 8\pi G\simeq 2\times
10^{-29}g$ cm$^{-3}$, close to that required for closure. The unit mass
involved is
\begin{equation}
m_{0}=\frac{h}{cL}=\frac{h}{c}\left( \frac{\Lambda }{3}\right) ^{1/2}\simeq
\,2\times 10^{-65}g\;\;.
\end{equation}
This is too small to be detected using current techniques and explains why
mass does not appear to be quantized.

The mass unit (10) is tiny even by the standards of particle physics, and
before proceeding to a presentation of more technical results a few comments
on concepts may be useful. \ The mass (10) of order 10$^{-65}$ g follows
from the length scale $\Lambda ^{-1/2}$ of order 10$^{28}$cm, or
equivalently the time scale of order 10$^{18}$s which is the age of the
universe. \ A more detailed analysis might alter the numbers somewhat (such
an analysis might, for example, involve the size of the particle horizon at
the current epoch or the current size of the cosmological ``constant'' in
models where this parameter varies). \ But the magnitude of (10) is based on
astrophysical data [32-36], and is expected to be correct to order of
magnitude. \ However, the interaction of a particle with mass $m_{o}$ of
order (10) with a vacuum of energy density of order $\Lambda c^{4}\diagup
8\pi G$ involves poorly-understood physics. \ Preliminary discussions of
this and related issues have been given recently [37, 38]. \ Our view, based
on (1) and the more general line element (13) below, is that a ``particle''
is just a localized concentration of energy in a medium where the
distinction between ordinary matter and the vacuum is convenient but
artificial. \ Energy, defined as the quantity which curves 4D space,
consists in general of contributions of both types [8, 37]. \ Indeed, energy
is a 4D concept that can be derived from 5D geometry, with Campbell's
theorem providing the link [21-24]. \ The small mass (10) simply reflects
the small (4D) curvature of the universe. \ In other problems, the parameter
$L$ in relations (1)-(10) can have different sizes. \ Consider the hydrogen
atom. \ (A detailed analysis of this problem is beyond the scope of the
present work, but would involve an exact solution of the 5D field equations
with electromagnetic potentials and orbital structure of the shell kind
discussed in reference 8, pp. 117-125.) \ The length scale is of order 10$%
^{-8}$ cm, or equivalently the time scale is of order 10$^{-18}$ s, so the
unit mass would be many orders different from that given by (10). \
Nevertheless, (10) defines the irreducible unit mass set by the energy
density of the background universe as measured by the cosmological constant.

The metric (1) which leads to (10) and the other results noted above involve
an algebraic choice for how $x^{4}=l$ enters that is suited to the physical
identification $l=h\diagup mc.$ Those results also depend on the assumption
that the 5D path is null. It is appropriate to call (1) the Plank gauge. To
put this gauge into the context of other work on 5D dynamics [1-20], let us
consider briefly how (1) may be altered in form and generalized.

Transforming (1) via $l\rightarrow L^{2}\diagup l$ gives
\begin{equation}
dS^{2}=\frac{l^{2}}{L^{2}}g_{\alpha \beta }\left( x^{\gamma }\right)
dx^{\alpha }dx^{\beta }-dl^{2},
\end{equation}%
which is pure canonical in form [2, 6, 8]. It is known that for (11) the 5D
field equations $R_{AB}=0$ contain the 4D Einstein equations in the form $%
G_{\alpha \beta }=3g_{\alpha \beta }\diagup L^{2}$, which describes a vacuum
spacetime with $G=-R=12\diagup L^{2}=4\Lambda $ so $\Lambda =3\diagup L^{2}$%
. (Here $R_{AB}$ is the 5D Ricci tensor, $G_{\alpha \beta }$ is the 4D
Einstein tensor and $R$ is the 4D Ricci scalar. The last relation was noted
above.) While it is a special case of Campbell's theorem, we see that any
solution of the 4D vacuum Einstein equations can be embedded in a solution
of the 5D vacuum field equations, including that of Schwarzschild [8]. This
implies that (11) is relevant\- to gravitational problems, and for this and
other reasons discussed below we will henceforth refer to (11) as the
Einstein gauge. It has been studied by several workers in the general case
where $g_{\alpha \beta }=g_{\alpha \beta }\left( x^{\gamma },\,l\right) $
and $dS^{2}\neq 0$ [2-8]. As mentioned before, we expect that this will lead
to violations of the Weak Equivalence Principle, since in general the
4-accelerations of test particles will depend on $l$, which whatever its
physical meaning may not be the same for all of them. This is born out by
the 5D geodesic equation, whose components yield equations of motion in
spacetime and the extra dimension which can be written thus:
\begin{subequations}
\begin{eqnarray}
\frac{du^{\mu }}{ds}+\Gamma _{\beta \gamma }^{\mu }\,u^{\beta }\,u^{\gamma }
&=&f^{\mu } \\[0.25in]
f^{\mu } &\equiv &\left( -g^{\mu \alpha }+\frac{1}{2}\frac{dx^{\mu }}{ds}%
\frac{dx^{\alpha }}{ds}\right) \frac{dx^{\beta }}{ds}\frac{dl}{ds}\frac{%
\partial g_{\alpha \beta }}{\partial l} \\[0.25in]
\frac{d^{2}l}{ds^{2}}-\frac{2}{l}\left( \frac{dl}{ds}\right) ^{2}+\frac{l}{%
L^{2}} &=&-\frac{1}{2}\left[ \frac{l^{2}}{L^{2}}-\left( \frac{dl}{ds}\right)
^{2}\right] \frac{dx^{\alpha }}{ds}\frac{dx^{\beta }}{ds}\frac{\partial
g_{\alpha \beta }}{\partial l}\;.
\end{eqnarray}%
Here $\Gamma _{\beta \gamma }^{\mu }$ are the usual 4D Christoffel symbols,
and $f^{\mu }$ is the fifth force (per unit inertial mass) which was first
found explicitly in the context of induced-matter theory [2] but also
figures in brane theory [9, 10]. Equation (12c) is second order, and is
identically satisfied with no constraint on $\partial g_{\alpha \beta
}\diagup \partial l$ by $l=l_{0}\exp \left[ \pm \left( s-s_{0}\right)
\diagup L\right] $, where $l_{0}$ and $s_{0}$ are arbitrary constants. This
is essentially (4) again, and by (11) means $dS^{2}=0$. [It can be verified
also that the reverse transformation $l\rightarrow L^{2}\diagup l$ which
converts (11) to (1) leaves the forms of (12a) and (12b) unchanged, while
(12c) retains its form for the r.h.s. and becomes $\left( d^{2}l\diagup
ds^{2}-l\,\,\diagup L^{2}\right) $ for the l.h.s.] We see that the Einstein
gauge (11) and the Planck gauge (1) describe essentially the same thing from
a mathematical perspective.

There is, however, an important difference from the physical perspective.
For the Planck gauge (1), we have argued that massive particles move on null
5D paths where the extra coordinate is $l_{P}=h\diagup mc$, a view which is
consistent with induced-matter theory [16] and equivalent by (4) with the
relation between the mass and the extra component of the momentum in brane
theory [20]. For the Einstein gauge (11), however, the relevant physical
identification is obviously $l_{E}=Gm\diagup c^{2}$. This parametization
goes back to at least 1990 [39], continues to the used [14], and is
consistent with the widely-held view that the essential role of fundamental
constants is to transpose physical dimensions [40, 41]. But it is important
to realize that it is \underline{only} in the Planck and Einstein gauges
that such simple parametizations of the mass hold. To appreciate why this is
so, reconsider (1) and (11). In the former, the proper distance in the fifth
dimension, defined in analogy with the proper time in general relativity, is
$\left| \int \left( L^{2}\diagup l^{2}\right) dl\right| =L^{2}\diagup
l\thicksim m$. In the latter, the proper distance is just $l\thicksim m$.
But in the general case the appropriate quantity to consider is $\int \left|
\epsilon g_{44}^{1/2}\right| dl=\int \left| \epsilon \Phi \right| dl$ where $%
g_{44}=\epsilon \Phi ^{2}\left( x^{A}\right) $ is the scalar potential [42,
43]. This is suppressed in (1), (11) but is generally present in 5D metrics
which have spacelike $\left( \epsilon =-1\right) $ or timelike $\left(
\epsilon =+1\right) $ extra dimensions. The appropriate line element to
replace (1), (11) is
\end{subequations}
\begin{equation}
dS^{2}=g_{\alpha \beta }\left( x^{\gamma },l\right) dx^{\alpha }dx^{\beta
}+\epsilon \Phi ^{2}\left( x^{\gamma },l\right) dl^{2}\;\;.
\end{equation}%
This is general insofar as it lacks $g_{4\alpha }$ but retains $g_{44}$ and
does not $l-$factorize $g_{\alpha \beta }$. The components of the 5D Ricci
tensor $R_{AB}$ for (13) can be worked out using tedious algebra (see ref.
8, p. 58). These can be applied to both induced-matter theory and brane
theory. For $R_{AB}=0$, the implied 15 relations yield naturally a set of 10
equations involving the 4D Einstein tensor, a set of 4 conservation
equations and 1 wave equation. The last is
\begin{equation}
\epsilon \,\Phi \,\square \,\Phi =-\frac{g_{\;\,\;\,,4}^{\lambda \beta
}\,g_{\lambda \beta ,4}}{4}-\frac{g^{\lambda \beta }\,g_{\lambda \beta ,44}}{%
2}+\frac{\Phi _{,4}\,g^{\lambda \beta }g_{\lambda \beta ,4}}{2\Phi }\;\;.
\end{equation}%
Here $\square \Phi $ $\equiv g^{\mu \nu }\Phi _{,\mu ;\nu }$ where a comma
denotes the partial derivative and a semicolon denotes the 4D covariant
derivative. Relation (14) has no analog in 4D since it comes from $R_{44}=0,$
and has been suggested in certain versions of 5D theory as being the
equation for the Higgs field which determines the masses of particles [8].
Another relation which follows from (13) for $R_{AB}=0$ and deserves
attention is that for the 4D Ricci scalar. This may be shown to be given by
\begin{equation}
R=\frac{\epsilon }{4\Phi ^{2}}\left[ g_{\;\,\,\,,4}^{\mu \nu }g_{\mu \nu
,4}+\left( g^{\mu \nu }g_{\mu \nu ,4}\right) ^{2}\right] \;\;.
\end{equation}%
This alters the status of $\Lambda $ as determined by vacuum spacetimes (see
above). Relations (14) and (15) require detailed study in regard to the
hierarchy problem and the cosmological-constant problem encountered by old
Kaluza-Klein theory.

As a last comment on how (1) may be altered in form and generalized, we
remark that the signature may be changed. This results in a two-time metric
of the type we have mentioned but ignored [25-30]. However, they have
interesting properties. For example, the first such found represents an
exact solution of the 5D field equations which describes a wave propagating
through a 4D de Sitter vacuum [25]. It is germane to point out that
wave-like behaviour cannot in general be obtained simply by applying a Wick
rotation to the fifth dimension as done in the Euclidean approach to 4D
quantum gravity [44]. Thus $l\rightarrow il$ in (1) or (2) with $dS^{2}=0$
just gives back $l=l_{0}\exp \left( \pm s\diagup L\right) $ as in (4), which
is a growing or shrinking mode. [The change is slow for $s\diagup L\ll 1$,
as pointed out above; and by (12) is dynamically undetectable anyway in the
pure Planck and Einstein gauges with $\partial g_{\alpha \beta }\diagup
\partial l=0$, since $f^{\mu }=0$ and the motion is geodesic in 4D.] Neither
are (1) and (2) altered by the so-called $\mathbf{Z}_{2}$ transformation $%
l\rightarrow -l$ of brane theory [20]. But if we take (1) or (2) with the
opposite sign for the last part of the metric and $dS^{2}=0$, there results $%
l=l_{0}\exp \left( \pm \,i\,s\diagup L\right) $. This is an oscillating
mode, and such deserve further study to see if they are related to the wave
nature of particles.

It is clear from the comments above that the Planck gauge (1) and the
Einstein gauge (11) are mathematically equivalent and physically special. So
why are they efficacious? The answer is that both have 4D parts which are
effectively momentum manifolds rather than coordinate manifolds. This is
achieved respectively through the mass parametizations
\begin{equation}
l_{P}=\frac{h}{mc},\,\;\;\;\;\,l_{E}=\frac{Gm}{c^{2}}\;\;.
\end{equation}%
These choices in old 4D theories of the scalar-tensor and scale-invariant
types were referred to as reflecting the use of atomic and gravitational
\underline{units} [45]; but in the present approach they refer to the use of
\underline{coordinates} in an underlying theory which is 5D covariant. In
the present approach, the Planck mass $\left( hc\diagup G\right) ^{\frac{1}{2%
}}$ as so defined lacks physical meaning, because it is a combination of
constants from both gauges (16) whose only purpose in either is to transpose
dimensions [40, 41, 45]. Put another way, the ratio $l_{E}\diagup
l_{P}=Gm^{2}\diagup ch$ can be formed and set equal to unity to produce $%
m=\left( ch\diagup G\right) ^{\frac{1}{2}}$, but this involves mixing
coordinates and is therefore badly defined.

\bigskip

\section{\protect\underline{Conclusion}}

There have been two recent results in five-dimensional relativity which are
remarkable. One is the existence in general of a fifth force which acts
parallel to the velocity in spacetime and can be related to a change in the
(inertial) rest mass of a particle. The other is the realization that a null
path in five dimensions can correspond to a timelike path in spacetime for a
massive particle. Both of these results require for their evaluation a
choice of coordinates or gauge. The Planck gauge (1) is so called because it
leads to the usual rule of quantization (6) and an uncertainty-type relation
(9). If the scale of the 5D geometry is related to the 4D cosmological
constant, there is a quantum of mass (10). The Einstein gauge (1) is so
called because it embeds the Schwarzschild solution, recognizes the Weak
Equivalence Principle as a symmetry of the metric and gives back 4D geodesic
motion. Both gauges are mathematically special but physically convenient
because their 4D parts effectively describe momentum manifolds rather than
coordinate manifolds. Both can be generalized, notably to (13) which applies
to all problems which do not involve explicit electromagnetic-type
potentials. This leads to a wave equation (14) for the scalar or Higgs
potential which has implications for the masses of particles, and an
embedding equation (15) for the scalar curvature of spacetime which has
implications for the size of the cosmological constant. While the Planck and
Einstein gauges (16) are well defined and the underlying theory is
covariant, a mixture of the two that produces the Planck mass is ill
defined, suggesting that if this parameter has meaning it does so in the
context of an $N\left( >5\right) D$ theory.

The current versions of 5D relativity that attract most attention are
induced-matter theory and membrane theory. However, it is worth recalling
that any 5D theory of the Kaluza-Klein type describes a spin-2 graviton, a
spin-1 photon and a spin-0 scalaron. While many topics for future work are
suggested by the outline given above, an obvious question concerns the
status of spin-%
$\frac12$
\ particles like the electron. Logically, an extension of the present
approach would give a geometrical account not only of bosons but also of
fermions. The usual approach to this, following Dirac, is of course to
factorize the 4D metric. But it is well known that in relativity the product
of the classical spin and path vectors of a particle can be made to vanish.
Preliminary work on this condition in 5D shows that it leads to a relation
that resembles the Dirac equation in 4D, with the mass entering as a
coordinate. A related question concerns how to relate the geometrical
approach to mass outlined above to the one in quantum field theory, where it
arises as the eigenvalue of the mass operator in the irreducible
representation of the appropriate symmetry group. If mass is geometrical in
nature and described by an $N\left( \geq 5\right) D$ field theory, the
latter will in general require the introduction of symmetry groups to
explain the masses and other properties of the observed elementary
particles. We hope that this and other problems will provide interesting
exercises for the reader.

\bigskip

\underline{{\Huge Acknowledgements}}

\bigskip

Thanks for comments go to H. Liu and S.S. Seahra. This work was supported by
N.S.E.R.C.

\bigskip

\underline{{\Huge References}}

\begin{enumerate}
\item Y.M Cho, D.H. Park, Gen. Rel. Grav. \underline{23} (1991) 741.

\item B. Mashhoon, H. Liu, P.S. Wesson, Phys. Lett. B \underline{331} (1994)
305.

\item P.S. Wesson, J. Ponce de Leon, Astron. Astrophys. \underline{294}
(1995) 1.

\item H. Liu, B. Mashhoon, Ann. Phys. (Leipzig) \underline{4} (1995) 565.

\item P.S. Wesson, B. Mashhoon, H. Liu, Mod. Phys. Lett. A \underline{12}
(1997) 2309.

\item B. Mashhoon, H. Liu, P.S. Wesson, Gen. Rel. Grav. \underline{30}
(1998) 555.

\item P.S. Wesson, B. Mashhoon, H. Liu, W.N. Sajko, Phys. Lett. B \underline{%
456} (1999) 34.

\item P.S. Wesson, Space-Time-Matter, World Scientific, Singapore (1999).

\item D. Youm, Phys. Rev. D \underline{62} (2000) 084002.

\item R. Maartens, Phys. Rev. D \underline{62} (2000) 084023.

\item A. Chamblin, Class. Quant. Grav. \underline{18} (2001) L17.

\item H. Liu, B. Mashhoon, Phys. Lett. A \underline{272} (2000) 26.

\item A. Billiard, W.N. Sajko, Gen. Rel. Grav. \underline{33} (2001) 1929.

\item W.B. Belayev, gr-qc/0110099 (2001).

\item J. Ponce de Leon, Phys. Lett. B \underline{523} (2001) 311.

\item P.S. Wesson, J. Math. Phys. in press (2002) (gr-gc/0105059).

\item V. Fock, Zeit. Phys. (Leipzig) \underline{39} (1926) 226.

\item A. Davidson, D.A. Owen, Phys. Lett. B \underline{177} (1986) 77.

\item S.S. Seahra, P.S. Wesson, Gen Rel. Grav. \underline{33} (2001) 1731.

\item D. Youm, Mod. Phys. Lett. A \underline{16} (2001) 2371.

\item J.E. Campbell, A Course of Differential Geometry, Clarendon, Oxford
(1926).

\item S. Rippl, C. Romero, R. Tavakol, Class. Quant. Grav. \underline{12}
(1995) 2411.

\item C. Romero, R. Tavakol, R. Zalaletdinov, Gen. Rel. Grav. \underline{28}
(1996) 365.

\item J.E. Lidsey, C. Romero, R. Tavakol, S. Rippl, Class. Quant. Grav.
\underline{14} (1997) 865.

\item A. Billyard, P.S. Wesson, Gen. Rel.\ Grav. \underline{28} (1996) 129.

\item A. Billyard, P.S. Wesson, Phys. Rev. D \underline{53} (1996) 731.

\item I. Bars, C. Kounnas, Phys. Rev. D \underline{56} (1997)\ 3664.

\item I. Bars, C. Deliduman, O. Andreev, Phys. Rev. D \underline{58} (1998),
066004.

\item I. Bars, C. Deliduman, D. Minic, Phys. Rev. D \underline{59} (1999)
125004.

\item J. Kocinski, M. Wierzbicki, gr-qc/0110075 (2001).

\item J.D. Bekenstein, Phys. Rev. D \underline{15} (1977) 1458.

\item C.H. Lineweaver, Astrophys. J. \underline{505} (1998) L69.

\item J.M. Overduin, F.I. Cooperstock, Phys. Rev. D \underline{58} (1998)
043506.

\item J.M. Overduin, Astrophys. J. \underline{517} (1999) L1.

\item M. Chiba, Y. Yoshii, Astrophys. J. \underline{510} (1999) 42.

\item J.M. Eppley, R.B. Partridge, Astrophys. J. \underline{538} (2000) 489.

\item P.S. Wesson, H. Liu, Int. J. Mod. Phys D \underline{10} (2001)\ 905.

\item F. Mansouri, hep-th/0203150 (2002).

\item P.S. Wesson, Gen. Rel. Grav. \underline{22} (1990) 707.

\item P.S. Wesson, Space Science Rev. \underline{59} (1992) 365.

\item J.W.G. Wignall, Phys. Rev. Lett. \underline{68} (1992) 5.

\item G.W. Ma, Phys. Lett. A \underline{143} (1990) 183

\item G.W. Ma, Phys. Lett. A \underline{146} (1990) 375.

\item G.W. Gibbons, S.W. Hawking (eds.), Euclidean Quantum Gravity, World
Scientific, Singapore (1993).

\item P.S. Wesson, Cosmology and Geophysics, Hilger/Oxford U. Press, New
York (1978).
\end{enumerate}

\end{document}